\documentstyle[12pt]{article}

 \newcommand{\be}{\begin{equation}}
 \newcommand{\ee}{\end{equation}}
 \newcommand{\ba}{\begin{eqnarray}}
 \newcommand{\ea}{\end{eqnarray}}
 
 \newcommand{\del}{\partial}

\newcommand{\lef}{\left}

\newcommand{\ri}{\right}

\newcommand{\cf}{{\cal F}}

\newcommand{\cl}{{\cal L}}

\newcommand{\fr}{\frac}

\begin{document}

\begin{titlepage}

\topmargin -15mm

\vskip 10mm

\centerline{ \LARGE\bf Quantum Global Strings}
\vskip 2mm
\centerline{ \LARGE\bf and Their Correlation Functions}

    \vskip 2.0cm

\centerline{\sc H. Fort$^1$ and E.C.Marino$^2$ }

\vskip 0.6cm

\centerline{$^1${\it Instituto de F\'\i sica, Facultad de Ciencias}}
\centerline{\it Universidad de la Rep\'ublica Oriental del Uruguay}
\centerline{\it Igu\'a 4225, 11400 Montevideo}  
\centerline{\it Uruguay}

\vspace{4mm}

\centerline{$^2${\it Instituto de F\'\i sica }}
\centerline{\it Universidade Federal do Rio de Janeiro }
\centerline{\it Cx.P. 68528, Rio de Janeiro RJ 21945-970}
\centerline{\it Brazil}

\vskip 1.0cm

\begin{abstract} 
 
A full quantum description of global vortex strings is presented
in the framework of a pure Higgs system with a
broken global U(1) symmetry in
3+1D. An explicit expression for the string creation operator is
obtained, both in terms of the Higgs field and in the dual formulation
where a Kalb-Ramond antisymmetric tensor gauge field is employed as the
basic field. The quantum string correlation function is evaluated and
from this, the string energy density is obtained. Potential
application in cosmology (cosmic strings) and condensed matter
(vortices in superfluids) are discussed.

\end{abstract}

\vskip 3cm
Work supported in part by Centro Latinoamericano de F\'\i sica (CLAF),
CNPq-Brazilian National Research Council and CSIC-Uruguayan Scientific
Research Council.


\end{titlepage}

\hoffset= -10mm

\leftmargin 23mm

\topmargin -8mm
\hsize 153mm
 
\baselineskip 7mm
\setcounter{page}{2}

\section{Introduction}

Quantized vortex lines or string-like excitations appear in a wide variety of
systems, ranging from condensed matter (vortices in Helium II) to cosmology 
(cosmic strings). In the case of superfluid He, there is still no 
microscopic theory which can explain several relevant issues such as 
the dynamics of quantum vortices, the 
dragging processes, the vortex nucleation or even the
structure of quantized vortex lines \cite{donnelly}.
On the other hand,
among the topological defects that were left behind by cosmological
phase transitions as the universe expanded and cooled,  
cosmic strings show very intersting properties. 
For instance, they may have
played an important role in the structure formation by acting as seeds of 
galaxies and other structures which can be observed today
in the universe \cite{ki}.

In all the standard approaches only classical
or semiclassical strings have been
considered. In a recent paper \cite{m96a} a full quantum theory of 
local (magnetic) strings in the Abelian-Higgs model was introduced.
In this work we present a fully quantized
formulation for strings or vortices in
a theory with a spontaneously broken {\it global} U(1) symmetry. Applying
the same strategy of refs.\cite{m96a}-\cite{mon}
we construct the corresponding quantum creation operator whose 
correlation functions are local (i.e. they
only depend on the positions of the string excitations).

The paper is organized as follows. In sec. 2 we consider the
simplest model which supports global
string-like excitations i.e. a complex scalar field with a symmetry breaking
potential and present a dual formulation
in terms of the Kalb-Ramond (K-R) antisymmetric tensor potential.
In sec. 3 we introduce the operator, $\sigma(C,t)$, that creates quantum 
topological strings along the curve $C$, which in the K-R language are
``charge'' strings. 
In sec. 4 we compute the correlation functions of this operator both
in the Kalb-Ramond and Higgs formulations.
In particular, we obtain the string tension for a long straight 
string.
Conclusions and final remarks are
presented in Section 5.

\section{Global Strings in the Higgs and Kalb-Ramond 
Representations}

Global string excitations appear in theories with a spontaneously broken 
continuous global symmetry.
We shall consider the simplest theory exhibiting 
string solutions, namely, that of a complex scalar field $\phi(x)$ described 
by the lagrangian density
\be
\cl[\phi] = |\partial_\mu \phi|^2 - V(|\phi|^2),
\label{eq:Lhiggs}
\ee
which has a global U(1) symmetry. The potential is the standard symmetry
breaking one:

$$V(|\phi|^2) = - m^2 |\phi|^2 + \fr{\lambda}{4} |\phi|^4$$.

Using the polar representation for the complex scalar field
$\phi = \fr{\rho}
{\sqrt{2}}e^{i\theta}$, the lagrangian density (\ref{eq:Lhiggs})
can be written as
\be
\cl[\rho,\theta] = \frac{1}{2} (\partial_\mu \rho)^2 +
\frac{1}{2}  \rho^2 (\partial_\mu \theta)^2 - V(\rho^2).
\label{eq:Lhiggs2}
\ee
The global U(1) invariance of (\ref{eq:Lhiggs2}) implies the
conservation of the current
\be
j_\mu=\rho^2\partial_\mu\theta = 
-i \phi^* \stackrel{\leftrightarrow}{\del}_\mu \phi.
\label{eq:J}
\ee
Field configurations containing strings correspond to a multivalued 
Goldstone $\theta$-field i.e. are such that the value 
of $\theta$ is defined up to $2\pi$ times an integer. It is convenient to
split the $\theta$-field into two parts
\be
\theta(t,\mbox{{\bf x}})=\bar{\theta}(t,\mbox{{\bf x}})+
\alpha(t,\mbox{{\bf x}}),
\label{t}
\ee
where $\bar{\theta}$ describes a given configuration of vortices
and $\alpha$ corresponds to the single valued part.
For a configuration of $N$ vortices, the
antisymmetric tensor vortex current is given by \footnote{
Topological currents are denoted by capital letters while Noether 
currents are denoted by lower case letters.}:
$$
J^{\mu \nu}(x) \equiv \epsilon^{\mu \nu \rho \sigma}\partial_\rho
\partial_\sigma \theta(x)
= \epsilon^{\mu \nu \rho \sigma}\partial_\rho
\partial_\sigma \bar{\theta}(x) =
$$
\be
= \sum_{a=1}^N \gamma_a
\int d\tau d\sigma (\stackrel{.}{X}_a^\mu X_a^{'\nu}
-\stackrel{.}{X}_a^\nu X_a^{'\mu})\delta^{(4)}(x-X_a(\tau,\sigma)),
\label{eq:J2}
\ee
where the subindex $a$ labels the vortices, $\gamma_a$ 
is the quantized circulation or vorticity,
$X_a$ denotes the vortex
position and the dot and prime indicate, respectively, differentiation
with respect to $\tau$ and $\sigma$.
The integrals are taken over the universe surfaces of the strings.
In the case of a superfluid
$\gamma_a=n_a\frac{h}{M}$ with $n_a=\pm 1,\pm 2,...$, where $M$ is
the mass of the superfluid atoms.
Observe that acting on the multivalued $\theta$-field,
different components of the derivative operator no longer commute.
The tensor topological current
$J^{\mu \nu}$ is identically conserved:
\be
\partial_\mu J^{\mu \nu}\equiv 0.
\nonumber
\label{eq:cons2}
\ee
This is because $\partial_\sigma \bar{\theta}(x) $ is no longer
multivalued.
It is possible to work in the equivalent dual representation expressed
in terms of the 
two-index antisymmetric Kalb-Ramond tensor field
$B_{\mu \nu}(x)$ instead of the
Goldstone boson $\theta(x)$. 
The connection with this dual formulation 
is provided by
\be
\fr{1}{2}
\epsilon_{\mu\nu\alpha\beta}\partial^\nu B^{\alpha\beta}=
\rho^2\partial_\mu \theta,
\label{eq:dualconn}
\ee
which can also be written as

$$^*\!H_\mu=\rho^2 \partial_\mu \theta,$$
or
\be
H^{\mu\nu\alpha} = \epsilon^{\mu\nu\alpha\beta} 
(\rho^2 \partial_\beta \theta)
\label{hte}
\ee
where $^*\!H_\mu=\fr{1}{6}\epsilon_{\mu\nu\alpha\beta}H^{\nu\alpha\beta}$ 
is the dual of the Kalb-Ramond field strength,
$H^{\alpha\beta\gamma}=\partial^\alpha B^{\beta \gamma} +
\partial^\beta B^{\gamma \alpha} + \partial^\gamma B^{\alpha \beta}$.
The $\theta$ term in (\ref{eq:Lhiggs2}) can be expressed in terms of
a Kalb-Ramond field, by means  of the following Gaussian identity 
\be
\exp \lef \{ i\int d^4x \lef [\frac{1}{2}\rho^2(\partial_\mu\theta)^2 \ri ]
\ri \} =
\int [\rho^{-1}d^*\!H_\mu]\exp \lef \{ i\int d^4x 
\lef [ -\frac{1}{2\rho^2}(^*\!H_\mu)^2
+ ^*\!H_\mu\partial^\mu\bar{\theta}+^*\!H_\mu\partial^\mu\alpha \ri ] \ri \}.
\label{id1}
\ee

Note that we have split the $\theta$-field in two parts, according to
(\ref{t}). Integrating over the single valued $\alpha$ we get a 
$\delta(\partial_\mu^*\!H^\mu )$ which is
identically satisfied, according to the definition of $^*\!H^\mu$.

Substituting (\ref{id1}) in the vacuum functional corresponding to the
lagrangian (\ref{eq:Lhiggs2}), integrating by parts the
$^*\!H_\mu$ derivatives in the
$\bar\theta$-term
and using (\ref{eq:J2}), we obtain
the following dual lagrangian density
\be
\cl[\rho,B_{\mu \nu}]= \frac{1}{2} (\partial_\mu \rho)^2 
-V(|\rho|^2) +
\fr{1}{12 \rho^2 } H_{\mu\nu\alpha}^2 +
\fr{1}{2} B_{\mu\nu}J^{\mu \nu}.
\label{eq:dualL}
\ee
The quantum theory corresponding to the above lagrangian density
is well defined because $<\rho> \neq 0$  (spontaneously broken symmetry).
On the contrary, for a non-spontaneously broken theory, if 
$<\rho> = 0$ the division by zero in the kinetic $H^2$ term
would make the theory undefined. 
In the approximation where the field $\rho$ has a constant value $\rho_0$ 
, i.e. the large $\lambda$ limit, (\ref{eq:dualL}) reduces to
\be
\cl[B_{\mu \nu}]= 
\fr{1}{12 \rho_0^2 } H_{\mu\nu\alpha}^2 +
\fr{1}{2} B_{\mu\nu}J^{\mu \nu}.
\label{eq:dualL2}
\ee

The corresponding operator field equation is:
\be
\del_\alpha H^{\alpha\mu\nu} = 
\, \rho_0^2 J^{\mu\nu}.
\label{eq:J3}
\ee
The theory possesses an identically conserved topological current:
\be
J^\mu = \fr{1}{2} \epsilon^{\mu\nu\alpha\beta} \del_\nu  B_{\alpha\beta},
\label{eq:J4}
\ee
which from (\ref{eq:dualconn}) is identical to the
Noether current (\ref{eq:J}): 
$J^\mu \equiv  j^\mu$, i.e. the topological 
current of the antisymmetric field 
coincides with the electric current (\ref{eq:J}) of the scalar
field representation.
On the other hand, the reciprocal result also does hold:
the topological current of the
Goldstone field representation (\ref{eq:J2}) appears as the
``electric'' current in the dual representation according to (\ref{eq:J3}). 
We see that   
the electric and topological currents in the original
and dual representations are interchanged, as usual.

\section{The Vortex String Creation Operator }

\setcounter{equation}{0}

Let us introduce now the creation operator for a fully quantized
string state. We have seen in the previous section that the 
topological charge associated to
the the string excitations become ``electric'' charges
in the dual K-R formulation. Hence in this formulation we need 
a charge creation operator.
This kind of operator has been introduced in \cite{qedt,m97} and, 
for a closed string $C$ it is given by
\be
\sigma(C,t) = \exp \lef \{ \fr{1}{2} \int d^4 x 
H_{\alpha\mu\nu}\tilde{C}^{\alpha\mu\nu} \ri \},
\label{eq:s0}
\ee
where the 3-tensor external field $\tilde{C}^{\alpha\mu\nu}$ is of the form
$$\tilde{C}^{\alpha\mu\nu}=\partial^\alpha\tilde{C}^{\mu\nu}$$
with
\be
\tilde{C}^{\mu\nu}=ia\int_{S(C)} d^2 \xi^{\mu\nu}\fr{1}{-\Box}
(z-\xi). 
\label{eq:C}
\ee
In the above expressions, $a$ is an arbitrary real 
number and $S(C)$ is a space-like
surface bounded by the closed string at $C$.
$d^2 \xi_{ij}$ is the surface element of $S(C)$, the directions $i,j$ being
along the surface.

Substituting (\ref{eq:C}) in (\ref{eq:s0}) we get for the $\sigma(C)$
operator the expression of ref.(\cite{m97}):
\be
\sigma(C,t) = \exp \lef \{ \fr{ia}{2} \int_{S(C)} d^2 \xi_{\mu\nu}
\fr{\del_\alpha H^{\alpha\mu\nu}}{-\Box} \ri \}
\label{s1}
\ee
or
\be
\sigma(C,t) = \exp \lef \{ \fr{-ia}{2} \int_{S(C)} d^2 \xi_{ij}
B^{ij} + gauge\ terms \ri \}.
\label{s2}
\ee
The gauge terms in (\ref{s2}) guarantee the gauge
invariance of $\sigma$ which is explicit in (\ref{s1}).
Later on, it will become clear
that both the correlation functions and commutation rules of $\sigma$
are independent of the surface $S$: they just depend on $C$. 
The generalization for an open string is straightforward.
The operator
$\sigma(C)$ creates a string along the curve C. In order to
prove this, let us consider the topological charge operator along a surface
$R$ (i.e. an analogous operator to the magnetic flux operator of 
ref.\cite{m97}, namely,
\be
\Phi_R = \int_R d^2 x^i J^{i0} (\vec x,t)
\label{fir}
\ee
Let us evaluate the commutator $[\Phi_R, \sigma]$. To do this, let 
us observe that according to (\ref{eq:J3})
$$
J^{i0} = \del_j \Pi^{ji}
$$
where $\Pi^{ij}$ is the momentum canonically conjugate to the Kalb-Ramond
field $B^{ij}$, satisfying the equal-time commutator
$$
[B^{ij} (\vec x,t), \Pi^{kl} (\vec y,t)] = i (\delta^{ik} \delta^{jl} -
\delta^{il} \delta^{jk}) \delta^3 (\vec x - \vec y)
$$
Using the above relation, (\ref{s2}), the Baker-Hausdorff formula and 
Stokes' theorem, we immediately get
\be
\lef [\int_R d^2 x^i J^{i0} (\vec x,t), \sigma(C,t) \ri ] = a \sigma(C,t) 
\int_R d^2 x^i \oint_C d\xi^i 
\delta^3 (\vec x - \vec y)
\label{js}
\ee
The above integrals give $\pm 1$ whenever 
the curve C pierces the surface $R$ in
the positive or negative sense, respectively. Otherwise they vanish. 
Hence, we if we choose the ``magnetic'' flux surface and the string 
in such a way that it pierces the surface positively, we get
\be
[ \Phi_R, \sigma ] = a\  \sigma
\label{fis}
\ee
This shows that the $\sigma$ operator carries $a$ units of ``magnetic'' flux 
along the curve $C$ and indeed creates a topological string along this curve.

\section{Quantum Strings Correlation Functions }

\setcounter{equation}{0}

\subsection{Kalb-Ramond Representation}

In this section we compute the Euclidean correlation functions
of the operator  $\sigma$ introduced above in the Kalb-Ramond
representation.
Using the
expression (\ref{eq:s0}) for the $\sigma$ operator
and the lagrangian (\ref{eq:dualL2}), we can express the correlation
function in Euclidean space (in a completely analogous way as in ref.
\cite{qedt}) as
$$
<\sigma (C_x)\sigma ^{\dagger}(C_y)>=Z^{-1}
\int DB_{\mu \nu} \exp\left \{-\int d^{4}z \left [\fr{1}{12 \rho_0^2}
H^{\mu\nu\alpha} H_{\mu\nu\alpha} + \right. \right.
$$
\be
\left. \left.
+\fr{1}{6}\tilde{C}_{\mu \nu \alpha}(z;x,y)
H^{\mu \nu \alpha} +
{\cal L}_R
 \right ]  \right \},
\label{cf1}
\ee
where ${\cal L}_R$ is a
surface renormalization factor, to be determined
below, which ensures the locality of
this correlation function i.e. the fact that it depends
only on the string position, namely, on the border of the surface, $C$,
and not on the surface $S(C)$ itself.

We see that $<\sigma\sigma^{\dagger}>=e^{W[\tilde{C}_{\mu \nu}]}$ 
is the vacuum
functional in the presence of the external field $\tilde{C}_{\mu \nu}$.
This property of the correlation functions of $\sigma$
is common to all of the topological charge bearing
related operators
\cite{m96a}-\cite{mon} and
follows from the general fact that  topological charge carrying
operators are closely related to the
disorder variables of Statistical Mechanics \cite{kc}. Indeed,
treating these operators as disorder variables
one can demonstrate in general \cite{evora} that the
$\sigma$ operator correlation functions can be expressed in terms of
the coupling of the lagrangian field to an external field like
$\tilde{C}_{\mu \nu \alpha}$ as in (\ref{eq:s0}). 
It is not difficult to see, using (\ref{s2}) and (\ref{eq:J2}) that the
second term in the exponent in (\ref{cf1}) can be written, up to gauge terms,
as
\be
\fr{1}{2} \int d^4z J_{\mu\nu} B^{\mu\nu},
\label{jb}
\ee
if we choose $a=\gamma_a$ and $N=1$ in (\ref{eq:J2}).
Going back to (\ref{eq:dualL}) or
(\ref{eq:dualL2}), and comparing with the above expression, we immediately 
conclude that if we retain the $\bar\theta$ part in
(\ref{id1}) and just integrate over the single valued part $\alpha$, the 
functional thereby obtained is precisely the above string correlation 
function $<\sigma \sigma^\dagger>$. This would provide an alternative way   
for obtaining the string operator.

One can show in general \cite{evora} that the appropriate
renormalization factor consists of the
corresponding self-coupling of 
the external field.
Also here, we will see explicitly that the renormalization
counterterm
\be
{\cal L}_R=\fr{\rho_0^2}{6}\tilde{C}^{\mu \nu \alpha}
\tilde{C}_{\mu \nu \alpha}
\label{lct}
\ee
will absorb all the hypersurface dependence of the correlation
function, thereby making it completely local. Indeed, performing the
change of functional integration variable

\be
B_{\mu\nu} \rightarrow B_{\mu\nu} + \rho_0^2 \Omega_{\mu\nu}
\label{cv}
\ee
with
\be
\Omega_{\mu\nu} = \tilde{C}_{\mu\nu} (S') - \tilde{C}_{\mu\nu} (S)
\label{ome}
\ee
-- where $S'$ is an arbitrary surface also bounded by the curve $C$ --
in the integral (\ref{cf1}), and choosing ${\cal L}_R$ as 
given by (\ref{lct}), we conclude, after a straightforward calculation, that
$<\sigma \sigma^\dagger>(S) = <\sigma \sigma^\dagger> (S')$, thereby
establishing its surface invariance in general.

Let us now explicitly compute
the correlation function (\ref{cf1}), with the choice
(\ref{lct}) made for $\cl_R$.
Before performing the functional integration in (\ref{cf1}),
note that we can rewrite the linear term as \cite{qedt}
\be
\fr{1}{6} \int d^4z \tilde C_{\mu\nu\alpha}
H^{\mu\nu\alpha} =
\fr{1}{2} \int d^4z K_{\mu\nu} 
B^{\mu\nu},
\label{lin}
\ee
where
\be
K_{\mu\nu} = \fr{1}{2} \tilde{C}_{\alpha \beta \gamma}
F^{\alpha\beta\gamma}\ _{\mu\nu},
\label{be}
\ee
and
\be
F^{\alpha\beta\gamma}\ _{\mu\nu} =\del^\alpha \Delta^{\beta\gamma}
\ _{\mu\nu} + \del^\beta \Delta^{\gamma\alpha}\ _{\mu\nu}
+ \del^\gamma \Delta^{\alpha\beta}\ _{\mu\nu}.
\label{efe}
\ee
with
\be
\Delta^{\mu\nu\alpha\beta}=\delta^{\mu\alpha}\delta
^{\nu\beta}-\delta^{\mu\beta}\delta^{\nu\alpha}.
\label{del}
\ee
Inserting (\ref{lct}) and a gauge fixing term
\be
{\cal L}_{GFB}=-\fr{\xi}{8\rho_0^2} B_{\mu \nu}K^{\mu \nu \alpha \beta}
(-\Box)^{-1}B_{\alpha \beta}
\label{gf}
\ee
-- where $K^{\mu \nu \alpha \beta}=\partial^{\mu}
\partial^{\alpha}\delta^{\nu \beta}+\partial^{\nu}\partial^{\beta}
\delta^{\mu \alpha}-(\alpha \leftrightarrow \beta )$ and $\xi$ is
gauge fixing parameter --
in (\ref{cf1}), we can perform the quadratic integration
over $B_{\mu \nu}$ with the help of the euclidean propagator
of this field, namely
\be
D^{\mu\nu\alpha\beta}(x)=\fr{\rho_0^2}{4} \lef
[(-\Box)\Delta^{\mu\nu\alpha\beta}+
(1-\xi^{-1})K^{\mu\nu\alpha\beta} \ri ] \left ( \fr{1}{\Box} \right )^2
\label{pro}
\ee
The result is
\be
<\sigma(C_x)\sigma^\dagger(C_y)> =
\exp \left \{ \fr{1}{2}\int d^4z d^4z'K^{\mu\nu}(z)
K^{\alpha\beta}(z') D_{\mu\nu\alpha\beta}(z-z')\right.
\left. - S_R \right \}.
\label{cf3}
\ee
where $S_R$ is the action corresponding to the renormalization
counterterm $\cl_R$.
 We immediately see that only the first term of (\ref{pro})
contributes to (\ref{cf3}). In particular all the gauge dependence
disappears.
 This happens
because of the gauge invariant way in which the external field is
coupled in (~\ref{cf1}) which results in the form of $K_{\mu\nu}$
given by (\ref{be}). Using the identity  \cite{qedt}
\be
F^{\mu\nu\alpha}\ _{\sigma\tau}\Delta^{\sigma\tau\lambda\chi}
F^{\gamma\rho\beta}\
_{\lambda\chi} = -4 \epsilon^{\mu\nu\alpha\sigma}
\epsilon^{\gamma\rho\beta\lambda} [ -\Box \delta^{\sigma\lambda}
+ \del^\sigma \del^\lambda ]
\label{identidade}
\ee
and performing the $z$ and $z'$ integrals in (\ref{cf3}), we get
\vfill\eject
$$
<\sigma (C_x) \sigma ^{\dagger}(C_y)>=
\exp \left \{
 \fr{a^{2}\rho_0^2}{2}\sum_{i,j=1}^{2}\lambda_{i}\lambda_{j}
\int_{S_i(C)}d^{2}\xi_{\alpha \beta} \int_{S_j(C)}d^{2}\eta_{\mu \nu}
\right.
$$
\be
\left.
\partial_\gamma \partial_\lambda'
\frac{1}{4}\left [ \frac{1}{-\Box} \right ]^2
\epsilon^{\alpha \beta \gamma \sigma}\epsilon^{\mu \nu \lambda \rho}
\left [ \Box \delta^{\sigma \rho} +\partial^\sigma \partial^\rho \right ]
\lef [\frac{1}{-\Box}\ri ] - S_R 
\right \}.
\label{cf2}
\ee
Since we are evaluating the two-point function, actually in
(\ref{cf1}), $\tilde{C}_{\mu \nu \alpha}(z;x,y) =
\tilde{C}_{\mu \nu \alpha}(z;x) - \tilde{C}_{\mu \nu \alpha}(z;y)$. Hence,
in the above expression, $i,j=1,2$ correspond to $x,y$, respectively, and
$\lambda_1 \equiv +1$ and  $\lambda_2 \equiv -1$.

Only the $\delta^{\sigma \rho}$ term of the above expression gives
a nonzero contribution. Inserting the identity
\be
\epsilon^{\alpha \beta \gamma \rho}\epsilon^{\mu \nu \lambda \rho} =
\delta^{\lambda\gamma} \Delta^{\alpha\beta\mu\nu}
- \delta^{\lambda\beta} \Delta^{\alpha\gamma\mu\nu}
+ \delta^{\lambda\alpha} \Delta^{\beta\gamma\mu\nu}
\label{idep}
\ee
in (\ref{cf2}) we immediately see that the
first term above produces an expression which
is exactly canceled by the surface renormalization counterterm $S_R$.
Applying Stokes' theorem to the expression yielded by the
two remaning terms we obtain
$$
<\sigma (C_x) \sigma ^{\dagger}(C_y)>=
\lim_{m,\epsilon \rightarrow 0 }
\exp \left \{
 \fr{ - a^{2}\rho_0^2}{2}\sum_{i,j=1}^{2}\lambda_{i}\lambda_{j}
\oint_{C_i}d\xi^{\alpha} \oint_{C_j}d\eta^{\alpha}
\right.
$$
\be
\lef. \lef [ -\fr{1}{8\pi^2} \ln \mu [ |\xi - \eta| + |\epsilon| ] \ri ]
\right \}.
\label{cf4}
\ee
The expression between brackets is $\fr{1}{(-\Box)^2} \equiv
\cf^{-1}\lef [\fr{1}{k^4} \ri ]$ and $\mu$ and $\epsilon$
are respectively an infrared and an ultraviolet regulator. We see that
for the above topological charge conserving correlation function, all the
$\mu$-dependence is cancelled because $\sum_i \lambda_i =0$. The $\epsilon$
dependence associated with the self-interacting $i-i$ terms can be
eliminated by a multiplicative renormalization of the string field operator
$\sigma$. Expression (\ref{cf4}) is precisely the one found for the large
distance behavior of a magnetic vortex in the case of a local U(1) Higgs
theory \cite{m96a}. 
This result, which at first sight might seem 
surprising, is actually to be expected,
according to the following argument.
In ref. \cite{qedt} it was proved that Maxwell lagrangian
\be
{\cal L_M}
 = -\fr{1}{4} F_{\mu \nu}F^{\mu \nu}.
\label{lm}
\ee
can be formulated
as a particular sector of a Kalb-Ramond theory with lagrangian
$$
{\cal L}'
 = -\fr{1}{12} H_{\mu \nu \alpha}(-\Box)^{-1}H^{\mu \nu
\alpha}.
$$
In a similar way, it is not difficult to show \cite{m97} that the lagrangian
\be
{\cal L}
 = -\fr{1}{12} H_{\mu \nu \alpha}H^{\mu \nu \alpha},
\label{lh}
\ee
appearing in (\ref{eq:dualL2}), corresponds to
$$
{\cal L}
 = -\fr{1}{4} F_{\mu \nu}(-\Box)^{-1}F^{\mu \nu}.
$$
It turns out that
this lagrangian has precisely the form of a gauge invariant mass term 
which appears in the local U(1) Higgs theory of ref. \cite{m96a} and
which determines the large distance behavior of the vortex correlation
function in the local case.
This explains the coincidence of results for the quantum
vortices in both models.

Considering the case of a straight string along
the $z$-direction and piercing the $z=0$ plane at the point $(\vec x, 0)$,
the global string creation operator $\sigma(C,t)$ can be written as
$\sigma(\vec x,t)$. Starting from (\ref{cf4}) and following exactly the same
steps as in \cite{m96a}, we obtain, for a long straight string of length
$L$,
\be
<\sigma(\vec x)\sigma^\dagger(\vec y)>=
\exp \left
 \{-\fr{La^2\rho_0^2}{8\pi}|\vec x-\vec y|,
\right \}
\label{cf5}
\ee
>From this expression, one can infer that the string energy is given by
$E(L)=\fr{La^2\rho_0^2}{8\pi}$
which means that the string energy density $\epsilon = E(L)/L$
is given by the following expression:
\be
\epsilon=\fr{a^2\rho_0^2}{8\pi}.
\label{epsilon}
\ee

The large distance behavior
of the correlation function of a genuine quantum vortex seems
to be ``universally" governed - whether they are local or global -
by an exponential decay. This is not the case, of course,
in the symmetric phase of
the U(1) Higgs model, where we have $\rho_0=0$ and consequently $E=0$. 
Hence it follows that $<\sigma \sigma ^{\dagger}> \longrightarrow$ const
$\neq 0$
when $|\vec x-\vec y| \rightarrow \infty$ and the conclusion is
that there are no true physical vortex excitations in this phase.

\subsection{Higgs Representation}

Let us evaluate here, for the sake of completeness, the string correlation
function in the Higgs language. As a subproduct, we will obtain the
global string creation operator in terms of the scalar Higgs field $\phi$.
Let us start from the correlation function (\ref{cf1}), with the $\cl_R$
given by (\ref{lct}). Using the fact that $\del_\mu\  ^*\!H^\mu \equiv 0$,
we can write (\ref{cf1}) as
$$
<\sigma (C_x) \sigma ^{\dagger}(C_y)>=
\int D^*\!H^\mu D\theta \exp \left \{ - \int d^4z \lef [ \fr{1}{2\rho_0^2}
(^*\!H^\mu)^2 +  \del_\mu \theta ^*\!H^\mu +
\ri. \ri.
$$
\be
\lef. \lef.
  ^*\!H^\mu \tilde A_\mu +
\fr{\rho_0^2}{6}  \tilde C_{\mu\nu\alpha} \tilde C_{\mu\nu\alpha} \ri ]
\ri \}
\label{c1}
\ee
where
\be
\tilde{A}^\mu = \epsilon^{\mu\nu\alpha\beta} \tilde{C}_{\nu\alpha\beta}
\equiv \epsilon^{\mu\nu\alpha\beta}\del_\nu \tilde{C}_{\alpha\beta},
\label{amu}
\ee
with $\tilde{C}_{\alpha\beta}$ given by (\ref{eq:C}).
Integrating over $^*\!H^\mu$, we obtain
$$
<\sigma (C_x) \sigma ^{\dagger}(C_y)>=
\int D\theta \exp \left \{ - \int d^4z \lef [ \fr{1}{2}
\rho_0^2 \del_\mu \theta \del^\mu \theta + \rho_0^2 \del_\mu
\theta \tilde{A}^\mu +
\ri. \ri.
$$
\be
\lef. \lef. 
\fr{1}{2} \rho_0^2 \tilde{A}^\mu \tilde{A}_\mu +
\fr{\rho_0^2}{6}  \tilde C_{\mu\nu\alpha} \tilde C_{\mu\nu\alpha} \ri ]
\ri \}
\label{c2}
\ee
We immediately recognize in the above expression the minimal coupling to
the external field $\tilde{A}^\mu$. Generalizing for the case of an arbitrary
value of $\rho$, we can write
\be
<\sigma (C_x) \sigma ^{\dagger}(C_y)>=
\int D\phi D\phi^* \exp \left \{ - \int d^4z \lef [ 
| D_\mu \phi |^2 + V(\phi) + \cl_R \ri ] \ri \}
\label{c3}
\ee
where $D_\mu = \del_\mu + i \tilde A_\mu$ with $\tilde A_\mu$ given by
(\ref{amu}).
Here, of course
$\tilde A_\mu \equiv \tilde A_\mu(x) - \tilde A_\mu (y)$ in order to
describe the two-point function.

>From this expression we can infer the explicit form of the string creation
operator in the Higgs representation:
\be
\sigma(C,t) = \exp \lef \{ \fr{1}{2} \int d^4 x 
(-i \phi^* \stackrel{\leftrightarrow}{D}_\mu \phi) \tilde{A}^\mu \ri \},
\label{s00}
\ee

Let us now evaluate (\ref{c3}) or (\ref{c2})
in the polar representation of $\phi$,
using the approximation of constant $\rho = \rho_0$.
Performing the quadratic $\theta$ integration in (\ref{c2}), we get
\be
<\sigma (C_x) \sigma ^{\dagger}(C_y)> =
\exp \lef \{ \fr{\rho_0^2}{2} \int d^4z 
 \tilde A_\mu \lef[ \fr{(-\Box \delta^{\mu\nu} + \del^\mu \del^\nu)}
{-\Box } \ri ] \tilde A_\nu  - S_R \ri \}
\label{cf7}
\ee
where the first term above is the tird term in (\ref{c2}) which did not
participate in the $\theta$-integration.
Inserting the explicit form of $\tilde A_\mu$, (\ref{amu}) in (\ref{cf7}),
we immediately reobtain (\ref{cf2}). From here on the calculation is
identical as before and we again arrive at expression (\ref{cf5}) for the
string correlation function. This establishes the equivalence of the
Higgs and KR formulations for the quantum string correlation functions.

\section{Conclusions}

The present method allows a description of the quantum dynamics 
of vortices or strings in a theory with a spontaneously broken {\it global}
U(1) symmetry. Temperature can be easily introduced in the usual
way, as the string correlation functions are simply given by functionals
of given peculiar external field configurations \cite{evora}.
This would allow one to
treat in a unified way both thermal and quantum fluctuations.
The potential applications of the present formalism are many.
A finite-temperature version of this treatment, for instance, might be
useful to study the creation of topological defects in the course of 
cosmological phase transitions \cite{zu96} in the Kibble/Zurek scenario.
Another interesting field of application can be found in condensed matter.
For example, by introducing an external
Lorentz-noninvariant background field \cite{ds}
this approach can be applied to a system like a superfluid helium film where 
point-like quantum vortices are induced at zero temperature 
as quantum fluctuations. In particular, it seems specially suited for
the problem of vortex nucleation. 
In general, this approach 
works in the dilute gas approximation i.e. as long as the core radius is
negligible compared with the vortex separation.

\vspace{4mm}

\leftline{\Large\bf Acknowledgements} \bigskip

This work was supported in part by the {\it Centro Latinoamericano de 
F\'{\i}sica} (CLAF).
We are grateful to R.Gambini for valuable discussions. H.F wants to thank
CSIC project No 052 for financial support. E.M. thanks CNPq for partial
financial support.

\end{document}